\documentclass[%
 reprint,
 showpacs,preprintnumbers,
 amsmath,amssymb,
 aps,
 prb,
]{revtex4-2}
\usepackage{dcolumn}
\usepackage{subfigure}
\usepackage{graphicx,color}
\usepackage{mathrsfs}

\begin{document}

\title{Emergence of spin-orbit coupling among spin, atomic orbital, and Bloch dynamics in Janus double-transition-metal MXenes}

\author{Tetsuro Habe}
\affiliation{Department of Mechanical and Electrical Systems Engineering, Kyoto University of Advanced Science, Kyoto 615-8577, Japan}

\date{\today}

\begin{abstract}
We found a spin-orbit coupling to cause a simultaneous correlation among three degrees of freedom, the electronic spin, orbital, and Bloch dynamics in an investigation into the electronic structure of Janus double-transition-metal MXenes, Mo$_2$HfC$_2$OS and W$_2$HfC$_2$OS. 
In this paper, it is also revealed that the spin-orbit coupling causes a staggered spin configuration with a trigonal pattern around the $\Gamma$ point near the insulating gap.
We developed a reduced Hamiltonian describing the electronic states and show that the spin-orbit coupling cannot be equated with conventional forms for a single electron in solids, LS, Rashba, and Dresselhaus couplings, even in the approximation under the low-energy and small wave number condition. 
Because of the intrinsic shape of the conduction band, a trigonally alternating spin-momentum locking emerges with the spin axis perpendicular to the layer plane.
The theoretical analysis shows that these Janus materials can provide a platform for exploring the spin-related phenomena due to the trigonal spin-momentum locking other than Rashba and Dresselhaus types.
\end{abstract}

\maketitle
\section{Introduction}

Spin-orbit coupling is a key ingredient of modern condensed matter physics in the last couple of decades.\cite{Galitski2013,Soumyanarayanan2016}
Although it is given originally by the coupling between kinetic and spin angular momenta as a relativistic correction to the electronic Hamiltonian, the spin-orbit coupling have emerged with various forms correlating the spin and several kinetic degrees of freedom in condensed matter physics.\cite{Dresselhaus1955,Bychkov1984}
In two-dimensional electronic systems, several types of spin-orbit coupling have caused fascinating electronic states: quantum spin Hall insulators\cite{Murakami2003,Konig2007,Tang2017} and spin-valley locking states of transition-metal dichalcogenides,\cite{Xiao2012,Kormanyos2013}
in crystalline materials even without strong electron-electron interaction.
Especially for semiconductors, the mathematically simple representations of spin-orbit coupling directly characterize the spin properties of the two-dimensional materials: Russell-Saunders coupling so-called LS coupling,\cite{Russell1925} Rashba coupling\cite{Bychkov1984}, and Dresselhaus coupling.\cite{Dresselhaus1955}
The last two couplings are varied with the Bloch dynamics of electron, the wave number, and require the absence of spatial parity symmetry in the crystal structure.

\begin{figure}[htbp]
\begin{center}
 \includegraphics[width=80mm]{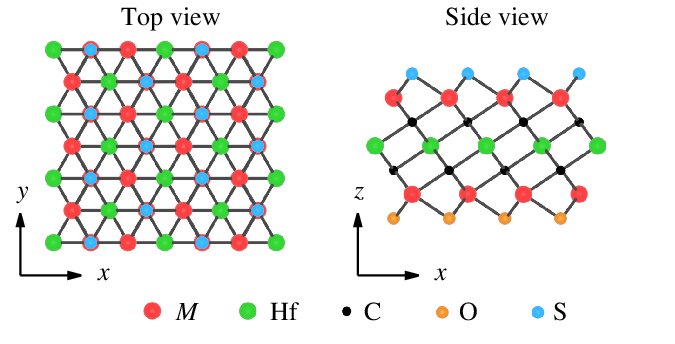}
\caption{The schematics of Janus double-transition-metal MXenes, $M_2$HfC$_2$OS for $M=$Mo and W. The left and right panels present the top view and side view of the monolayer.
The $z$-axis is implemented the perpendicular direction to the layer plane.
 }\label{fig_schematic}
\end{center}
\end{figure}

Mo$_2$HfC$_2$OS and W$_2$HfC$_2$OS, the target materials shown in Fig.\;\ref{fig_schematic}, are Janus monolayers, in which spatial parity symmetry is broken by different elements on two surfaces,\cite{Lu2017janus,Zhang2017janus} in the equivalent structure to a double-transition-metal MXene, Mo$_2$HfC$_2$O$_2$.\cite{Khazaei2016,Si2016,Liang2017}
In general, MXenes consist of alternately stacked sublayers of transition-metal elements and carbon or nitrogen element.\cite{Naguib2014,Gogotsi2019}
However, these monolayers are not synthesized as the pristine materials and they are terminated with other elements, e.g., fluorine, oxygen, hydroxy group, etc., corresponding to the fabrication condition and/or post-synthesis process.\cite{Lim2022,Thangavelu2026}
The Janus monolayers, $M_2$HfC$_2$OS for $M=$Mo and W, are given by replacing the oxygen terminating one surface of $M_2$HfC$_2$O$_2$ with sulfur in the same group of oxygen.
Although the symmetrically-terminated counterparts possess a large spin-orbit coupling constant, the effect of the coupling is an energy gap with no spin-split in the band structure because of spatial parity symmetry.\cite{Khazaei2016}
On the other hand, the two different terminations of $M_2$HfC$_2$OS break spatial parity symmetry in the crystal structure as shown in Fig.\;\ref{fig_schematic}, and it can produce a spin-split which is absent in the symmetric counter parts $M_2M'$CO$_2$ and $M_2M'$CS$_2$.


In this paper, we focus especially on the spin-orbit coupling and the spin configuration emerging in the Janus monolayers of double-transition-metal MXenes.
The intrinsically large coupling constant can be expected to enhance the spin-orbit coupling newly generated due to breaking spatial parity symmetry.
We develop a reduced theoretical model to describing the electronic states and show that the presence of spin-orbit coupling qualitatively inequivalet to conventional one: LS, Rashba, and Dresselhaus couplings.
Moreover, it is revealed that the unconventional spin-orbit coupling causes a staggered spin polarization with a trigonal pattern and the tilt of spin direction depending on the wave number with respect to the perpendicular axis to the layer plane.
Then, we show that the characteristic behavior of spin is attributed to the unconventional spin-orbit coupling emerging in the Janus materials.


The main body of this paper is organized as follows.
In Sec.\;\ref{sec_first-principles}, the electronic band structures are given using the first-principles calculations for the Janus double-transition-metal MXenes. 
The crystal structure is also optimized within the same numerical method. 
Then, the characteristics of spin-split are investigated to specify the property of the spin-orbit coupling in these materials.
In Sec.\;\ref{sec_effective_model}, an effective Hamiltonian is developed to describing three bands around the energy gap near the $\Gamma$ point.
The mathematical representations are acquired referring the orbital characters of the bands and the symmetries preserved in the materials.
Especially for the spin-orbit coupling, the reduced forms are also given within the conduction and valence bands.
In Sec.\;\ref{sec_spin_configuration}, the spin configuration in the bands is investigated using the effective model.
The discussion and conclusion are given in Sec.\;\ref{sec_discussion} and Sec.\;\ref{sec_conclusion}, respectively.


\section{First-principles Calculation}\label{sec_first-principles}


\begin{table}
\caption{The atomic positions in a unit cell for $M_2$HfC$_2$OS with two transition-metal elements $M$ and Hf. The parameters $(\xi_1,\xi_2,\xi_3)$ represents the position via $\boldsymbol{r}=\xi_1\boldsymbol{a}_1+\xi_2\boldsymbol{a}_2+\xi_3c\boldsymbol{e}_z$.
}
\begin{ruledtabular}
\begin{tabular}{c c c c c c c c}
&O&$M$(1)&C(1)&Hf&C(2)&$M$(2)&S\\ \hline
$\xi_1$&2/3&-2/3&2/3&0&-2/3&2/3&-2/3\\ 
$\xi_2$&1/3&-1/3&1/3&0&-1/3&1/3&-1/3\\ 
$\xi_3$&-1/2&$d_{M(1)}/c$&$d_{\mathrm{C}(1)}/c$&$d_{\mathrm{Hf}}/c$&$d_{\mathrm{C}(2)}/c$&$d_{M(2)}/c$&1/2\\ 
\end{tabular}\label{tab_atomic_position}
\end{ruledtabular}
\end{table}


The properties of spin-orbit coupling are represented by the spin-split in the electronic band structure obtained using first-principles calculations for Mo$_2$HfC$_2$OS and W$_2$HfC$_2$OS.
The two-dimensional lattice structure is depicted in Fig.\;\ref{fig_schematic} and represented by two basis vectors, $\boldsymbol{a}_1=(\sqrt{3}a/2,-a/2)$ and $\boldsymbol{a}_2=(0,a)$, with a single lattice parameter $a$.
The vertical thickness of the monolayer $c$ is defined by the distance between O and S.
The configuration of each element is represented by the ratios $\xi_j$ to the lattice vectors and the thickness as shown in table \ref{tab_atomic_position}.
The horizontal configuration of each element is the same as symmetrically-terminated double-transition-metal MXenes.\cite{Khazaei2016,Si2016,Huang2020,Parajuli2024}
However, the vertical position $d_\alpha$ of each sublayer is different from that in the symmetrically-terminated counterparts.
The middle sublayer of Hf is not placed at the center of a single layer because of the absence of spatial parity symmetry.
Thus, the crystal structure can be specified by the parameters, the lattice constant $a$, the thickness $c$, and $d_\alpha$.
The parameters are numerically estimated using density functional theory (DFT) as presented in table \ref{tab_lattice_parameters}.
The numerical calculation is performed using quantum-espresso, \cite{Quantum-espresso} a package of numerical codes for DFT, with Perdew-Burke-Ernzerhof functional\cite{PBE_functional} using the projector augmented wave method in PSlibrary (v1.0.0).\cite{Corso2014}
Under DFT calculations, monolayers are periodically aligned in the $z$ direction due to the periodic boundary condition.
Thus, to avoid the interaction between adjacent layers, the vacuum spacing of 40\AA\ is introduced between adjacent layers.
In the optimization, the criteria $2\times10^{-2}$ kbar and $10^{-4}$ Ry/Bohr are adopted for the pressure and force, respectively.
The energy convergence criterion is $10^{-8}$ Ry for self-consistent field calculations.
The energy cutoff is 70 Ry for the plane wave basis and 650 Ry for the charge density on the $12\times12\times1$ $k$-mesh in the Brillouin zone.


\begin{table}
\caption{The crystal parameters optimized by DFT for Mo$_2$HfC$_2$OS and W$_2$HfC$_2$OS. These parameters are presented in the unit of \AA.
}
\begin{ruledtabular}
\begin{tabular}{c c c c c c c c}
&a&c&$d_{M(1)}$&$d_{\mathrm{C}(1)}$&$d_{\mathrm{Hf}}$&$d_{\mathrm{C}(2)}$&$d_{M(2)}$\\ \hline
Mo&3.057&8.095&-2.883&-1.612&-0.226&1.162&2.412\\ 
W&3.053&8.116&-2.885&-1.617&-0.226&1.165&2.416\\ 
\end{tabular}\label{tab_lattice_parameters}
\end{ruledtabular}
\end{table}


\begin{figure}[htbp]
\begin{center}
 \includegraphics[width=80mm]{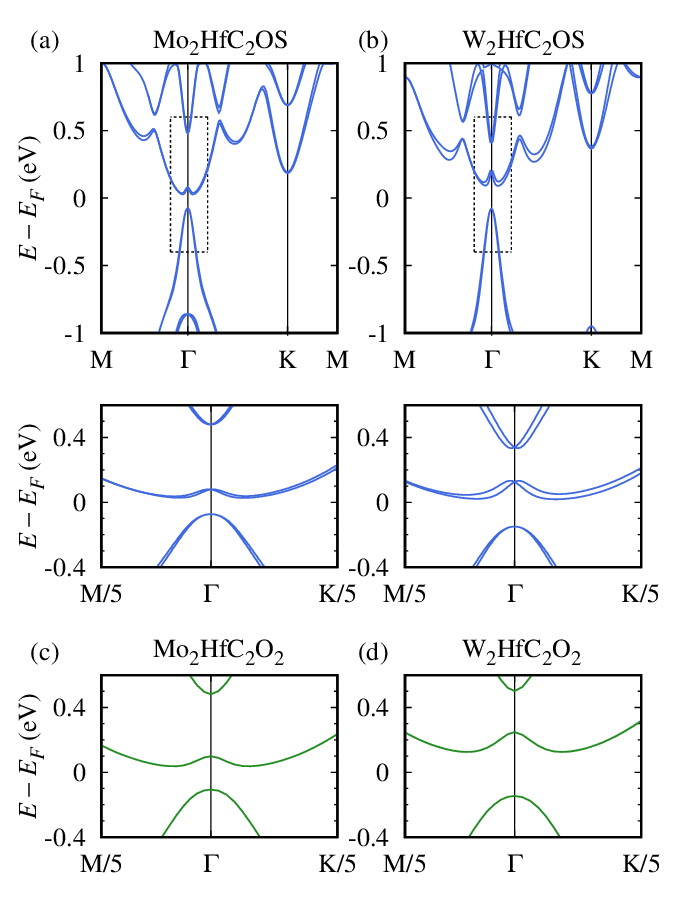}
\caption{The band structures of Janus MXene, $M_2$HfC$_2$OS, and inversion-symmetric MXenes, $M_2$HfC$_2$O$_2$. In (a) and (b), the bands of Mo$_2$HfC$_2$OS and W$_2$HfC$_2$OS are presented, respectively. The lower panels gives the dispersion within the boxes in the upper panels. In (c) and (d), those of Mo$_2$HfC$_2$O$_2$ and W$_2$HfC$_2$O$_2$ are presented, respectively.
 }\label{fig_bands}
\end{center}
\end{figure}


The spinful electronic band structures are presented in Fig.\;\ref{fig_bands} where the calculations are performed under the same conditions as the structural optimization on quantum-espresso.
Here, the crystal structures are represented by the parameters in table \ref{tab_lattice_parameters} and those for $M_2$HfC$_2$O$_2$ in Ref.\;\onlinecite{Habe2025-1,Habe2025-2}.
The spin-orbit coupling (SOC) is implemented adopting fully-relativistic functional for the calculations.
In (a) and (b), the upper and lower panels present the spinful band structures in the whole Brillouin zone and the vicinity of the $\Gamma$ point, 1/5 of each reciprocal space distance, for Mo$_2$HfC$_2$OS and W$_2$HfC$_2$OS.
The band structure indicates the presence of an energy gap which decreases near the $\Gamma$ point.
Each band splits into two branches due to the spin-orbit coupling associated with spatial parity symmetry breaking in the Janus-MXenes. 
For $M_2$HfC$_2$O$_2$, which preserves spatial parity symmetry, on the other hand, possesses the spin degeneracy in all bands in Fig.\;\ref{fig_bands} (c) and (d).
The spin-split of bands indicates the change of spin-orbit coupling qualitatively from that in the symmetrically-terminated MXenes.


In these materials, the properties of the spin-orbit coupling around the $\Gamma$ point is crucial because electronic states near the energy gap are relevant to the electronic transport, optical properties, and spin-related phenomena.
In the lower panels of Fig.\;\ref{fig_bands}(a) and (b), the behavior of the spin-split bands around the $\Gamma$ point are shown for Mo$_2$HfC$_2$OS and W$_2$HfC$_2$OS.
All bands possess the point degeneracy at the $\Gamma$ point and split into two spin branches with an increase in the wave number. 
In the spin-split around the $\Gamma$ point, some qualitative differences are observed in comparison with the conventional terms, Rashba coupling\cite{Bychkov1984} and Dresselhaus coupling.\cite{Dresselhaus1955}
Although two conduction bands exhibit the linearly crossing at the intersection of branches, like Rashba spin-orbit coupling, the valence band splits with no linear intersection.
Moreover, the spin-split is explicitly anisotropic around the $\Gamma$ point.
Especially in the lowest conduction band, the spin-split decreases along the $\Gamma$-M path, but it is retained along the $\Gamma$-K path.
The absence of a monotonic increase with the wave number is also the difference from the conventional types of spin-orbit coupling.
These features characterize the spin-orbit coupling generated by the asymmetric termination on the Janus MXenes.


\section{Analytic Representation of SOC}\label{sec_effective_model}

\begin{figure}[htbp]
\begin{center}
 \includegraphics[width=80mm]{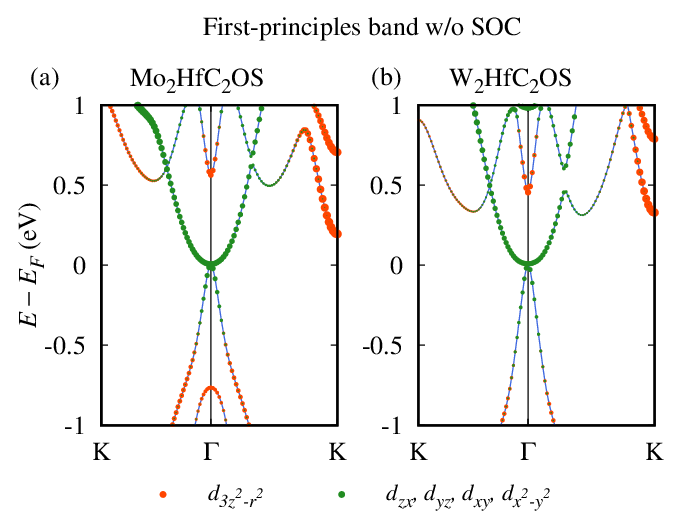}
\caption{The first-principles band structures in the absence of the spin-orbit coupling. In (a) and (b), the spinless bands of Mo$_2$HfC$_2$OS and W$_2$HfC$_2$OS are presented, respectively. The size of symbol represents the amplitude of atomic orbitals in transition-metal elements.
 }\label{fig_bands_wo_soc}
\end{center}
\end{figure}

 
In the vicinity of the energy gap, the description using an effective model is useful to investigate the properties of electronic states qualitatively.
In the case of symmetrically-terminated monolayers, $M_2$HfC$_2$O$_2$, these electronic states appear only around the $\Gamma$ point and they can be described using the three-band model proposed in Ref.\;\onlinecite{Habe2025-1,Habe2025-2}.
Without the spin-orbit coupling, the Janus monolayers, $M_2$HfC$_2$OS, also preserve the equivalent properties of electronic structure to the symmetrically-terminated counterparts as shown in Fig.\;\ref{fig_bands_wo_soc}.
Here, the amplitude of atomic orbitals in the transition-metal elements are also depicted as circular symbols.
The conduction and valence bands touch at the $\Gamma$ point and they include the $d$-orbitals with non-zero angular momenta on the point.
The first excited band, on the other hand, contains the $d_{3z^2-r^2}$ orbital only, which has zero angular momentum, on that point.
Moreover, $M_2$HfC$_2$OS preserves the same set of symmetries except for spatial parity symmetry.
Therefore, the three-band model can be available as the Hamiltonian for the Janus MXenes without the spin-orbit coupling.


\subsection{Spinless Hamiltonian}


In the present configuration, the three-band Hamiltonian is given by
\begin{align}
H_0({\boldsymbol{k}})=\begin{pmatrix}
E_1&-iv ke^{i\theta_{\boldsymbol{k}}}&-iv ke^{-i\theta_{\boldsymbol{k}}}\\
iv ke^{-i\theta_{\boldsymbol{k}}}&0&-u k^2e^{-2i\theta_{\boldsymbol{k}}}\\
iv ke^{i\theta_{\boldsymbol{k}}}&-u k^2e^{2i\theta_{\boldsymbol{k}}}&0
\end{pmatrix},\label{eq_spinless_Hamiltonian}
\end{align}
with the wave number $\boldsymbol{k}=(k\cos\theta_k, k\sin\theta_k)$ around the $\Gamma$ point $\boldsymbol{k}=0$.
Here, the parameters $E_1$, $u$, and $v$ represent the mixing of three composite orbitals isolated at the $\Gamma$ point.
At the $\Gamma$ point, the electronic states are the eigenstates of threefold rotation operator, and they are characterized by three eigenvalues, $R_3=1$, $\omega$, and $\omega^\ast$, with $\omega=\exp[2\pi i/3]$.
Since the $\Gamma$ point is also invariant under the mirror reflection along the $y$-axis, which exchanges $\omega$ and $\omega^\ast$, the two orbitals of $\omega$ and $\omega^\ast$ must be degenerate.
Therefore, the basis of three orbitals for Eq.\;(\ref{eq_spinless_Hamiltonian}) is represented by
\begin{align}
\Psi_{\Gamma}=\left(|1\rangle,\;|\omega\rangle,\;|\omega^\ast\rangle\right),\label{eq_basis}
\end{align}
with the eigenstate $|R_3\rangle$.
Here, $|1\rangle$ contains the $p$ and $d$ orbitals with zero angular momentum only.
In the case of $|\omega\rangle$ ($|\omega^ast\rangle$), the state consists of orbitals obtaining $\omega$ ($\omega^\ast$) under threefold rotation, e.g., $d_{zx}+id_{yz}$ and $d_{xy}-id_{x^2-y^2}$ ($d_{zx}-id_{yz}$ and $d_{xy}+id_{x^2-y^2}$).
Since the three states obtain different phases under threefold rotation, the mixing among them depends on $(k_x\pm ik_y)$, i.e., $\theta_{\boldsymbol{k}}$ at non-zero waver numbers due to the threefold rotational symmetry.
Moreover, the equivalence between $|\omega\rangle$ and $|\omega^\ast\rangle$ under the mirror operation requires the same coupling constant $v$ to the mixing with $|1\rangle$.
In contrast to the symmetrically-terminated counterpart, the basis does not preserve spatial parity symmetry.


The spinless model in Eq.\;(\ref{eq_spinless_Hamiltonian}) describes that the lowest conduction and highest valence bands are nearly isolated from the excited band around the $\Gamma$ point.
The effective Hamiltonian is analytically diagonalized,
\begin{align}
\hat{\Psi}_0^\dagger H_0(\boldsymbol{k})\hat{\Psi}_0=\mathrm{diag}\left[E_{\mathrm{ex}},\;E_{\mathrm{c}},\;E_{\mathrm{v}}\right],\label{eq_spinless_energy}
\end{align}
with the energy dispersion of each band,
\begin{align}
\begin{split}
E_{\mathrm{ex}}&=\frac{E_1-uk^2}{2}+\frac{E_1+uk^2}{2}\sqrt{1+r^2},\\
E_{\mathrm{c}}&=uk^2,\\
E_{\mathrm{v}}&=\frac{E_1-uk^2}{2}-\frac{E_1+uk^2}{2}\sqrt{1+r^2},
\end{split}
\end{align}
where the coefficient $r$ depends on the wave number $k$,
\begin{align}
 r=\frac{2\sqrt{2}v}{E_1+uk^2}k,
\end{align}
and it vanishes at the $\Gamma$ point.
Here, the unitary operator $\hat{\Psi}_0$ is given by
\begin{align}
\hat{\Psi}_0=\frac{1}{\sqrt{2}}\begin{pmatrix}
\sqrt{2}c_r&0&i\sqrt{2}d_r\\
id_re^{-i\theta_k}&-\sqrt{2}e^{-i\theta_k}&c_re^{-i\theta_k}\\
id_re^{i\theta_k}&\sqrt{2}e^{i\theta_k}&c_re^{i\theta_k}
\end{pmatrix},\label{eq_state_matrix}
\end{align}
with two coefficients,
\begin{align}
c_r=\sqrt{\frac{1}{2}+\frac{1}{2\sqrt{1+r^2}}},\;\;d_r=\frac{r}{\sqrt{2(1+r^2+\sqrt{1+r^2}})}.
\end{align}
Each column in Eq.\;(\ref{eq_state_matrix}) represents the eigenstate for the energy in Eq.\;(\ref{eq_spinless_energy}) and clearly shows the suppression of hybridization of $|1\rangle$ and others, i.e., $|\omega\rangle$ and $|\omega^\ast\rangle$, near the $\Gamma$ point, i.e., $d_r\ll1$.


\begin{figure}[htbp]
\begin{center}
 \includegraphics[width=80mm]{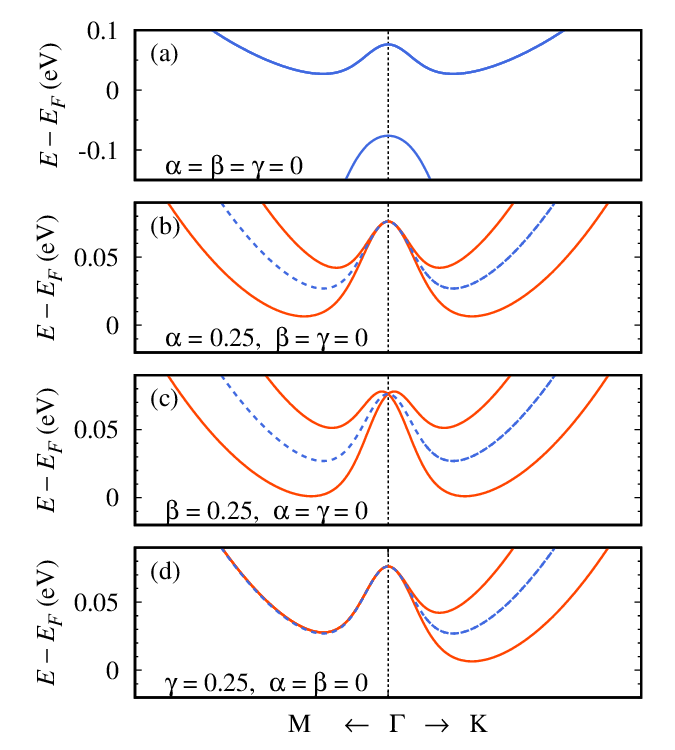}
\caption{The band structures obtained utilizing the effective model in Eq.\;(\ref{eq_spinless_Hamiltonian}) with the spin-orbit coupling. In (a), only the LS coupling is present. In (b)-(c), the band structure is presented by the solid line with one of the three terms, $H_{\mathrm{soc}}^{\mathrm{R}}$, $H_{\mathrm{soc}}^{xy}$, and $H_{\mathrm{soc}}^z$, switched on.
 }\label{fig_effective_bands}
\end{center}
\end{figure}


\subsection{Spin-orbit Coupling}


In the presence of spin-orbit coupling, the orbitals in the two degenerate bands possess a constant coupling so-called Russell-Saunders coupling, so-called LS coupling, $H_{\mathrm{soc}}^{\mathrm{LS}}$ because of the non-zero orbital angular momentum.
Since the crystal symmetry restricts the spin direction for this coupling to the $z$ axis, it is represented by
\begin{align}
H^{\mathrm{LS}}_{\mathrm{soc}}=-\lambda\begin{pmatrix}
0&0&0\\
0&\hat{\sigma}_z&0\\
0&0&-\hat{\sigma}_z
\end{pmatrix},\label{eq_LS_coupling}
\end{align}
with the Pauli matrix $\hat{\sigma}_z=\mathrm{diag}[1,-1]$ and the coupling constant $\lambda$ corresponding to each material.
This type of spin-orbit coupling is held regardless of spatial parity symmetry, i.e., it is also presented in symmetrically-terminated MXenes.
Near the $\Gamma$ point, i.e., $d_r\rightarrow0$ and $c_r\rightarrow1$, the LS coupling can be rewritten as the reduced form,
\begin{align}
h^{\mathrm{LS}}_{\mathrm{soc}}=\lambda\begin{pmatrix}
0&\hat{\sigma}_z\\
\hat{\sigma}_z&0\\
\end{pmatrix},\label{eq_LS_effective}
\end{align}
on the basis of the lowest conduction and highest valence bands $(|E_{\mathrm{c}}\rangle, |E_{\mathrm{v}}\rangle)$, the lower $2\times2$ block of $\hat{\Psi}_0^\dagger H_{\mathrm{LS}}\hat{\Psi}_0$.
The coupling generates the energy gap at the $\Gamma$ point, but it retains the double degeneration in each band as shown in Fig.\;\ref{fig_effective_bands} (a).


In the absence of spatial parity symmetry, three additional types are allowed as the spin-orbit coupling within the linear order of the wave number $k$.
They couples the spin and the other degrees of freedom: the atomic orbitals and the Bloch dynamics, i.e., the wave number.
The mathematical formula are restricted by time-reversal, threefold rotation, and mirror reflection symmetries on the basis in Eq.\;(\ref{eq_basis}) with the spin degree of freedom.
For time-reversal operation $\mathcal{T}$ and mirror reflection operation along the $y$-axis $\mathcal{M}_y$, the transformations for a spin-orbit coupling Hamiltonian $H_{\mathrm{soc}}(k_x,k_y)$ are defined as,
\begin{align}
\begin{split}
\mathcal{T}^{-1}H_{\mathrm{soc}}(k_x,k_y)\mathcal{T}=&(i\hat{\sigma}_yU)^\dagger H_{\mathrm{soc}}^\ast(-k_x,-k_y)(i\hat{\sigma}_yU),\\
\mathcal{M}_y^{-1}H_{\mathrm{soc}}(k_x,k_y)\mathcal{M}_y=&(i\hat{\sigma}_yU)^\dagger H_{\mathrm{soc}}(k_x,-k_y)(i\hat{\sigma}_yU),\label{eq_time_and_mirror}
\end{split}
\end{align}
with a unitary matrix exchanging $|\omega\rangle$ and $|\omega^\ast\rangle$,
\begin{align}
U=\begin{pmatrix}
1&0&0\\
0&0&1\\
0&1&0
\end{pmatrix},
\end{align}
on the spinless basis.
Moreover, the transformation of threefold rotation is preformed as follows
\begin{align}
\mathcal{R}_{1/3}^{-1}H_{\mathrm{soc}}(k_x,k_y)\mathcal{R}_{1/3}=\left(Ve^{-i\frac{\pi}{3}\hat{\sigma}_z}\right)^\dagger H_{\mathrm{soc}}(\tilde{k}_x,\tilde{k}_y)Ve^{-i\frac{\pi}{3}\hat{\sigma}_z},\label{eq_rotation}
\end{align}
with the phase shift,
\begin{align}
V=\mathrm{diag}[1,e^{-2\pi i/3},e^{2\pi i/3}],\label{eq_phase_shift}
\end{align}
and the rotated wave number $(\tilde{k}_x,\tilde{k}_y)$ with $2\pi/3$ in the two-dimensional space $(k_x,k_y)$.
Any additional types of coupling must be invariant under the three transformations, $\mathcal{T}$, $\mathcal{M}_y$, and $\mathcal{R}_{1/3}$.


Among the three additional types, one contains only diagonal parts on the orbital basis in Eq.\;(\ref{eq_basis}), and the others consist of off-diagonal parts, i.e., the inter-orbital coupling.  
The diagonal parts must be Rashba-type coupling under the restrictions in Eqs.\;(\ref{eq_time_and_mirror}) and (\ref{eq_rotation}).
Moreover, time-reversal symmetry requires the coupling to hold the same coefficient on the two orbitals $|\omega\rangle$ and $|\omega^\ast\rangle$,
\begin{align}
H^{\mathrm{R}}_{\mathrm{soc}}=-(k_x\hat{\sigma}_y-k_y\hat{\sigma}_x)\begin{pmatrix}
\alpha_0&0&0\\
0&\alpha&0\\
0&0&\alpha
\end{pmatrix}.
\end{align}
On the other hand, the inter-orbital coupling contains not only the in-plane spin $H_{\mathrm{soc}}^{xy}$ but also the out-of-plane spin $H_{\mathrm{soc}}^{z}$.
Since threefold rotation adds a phase factor to them, the components of $H_{\mathrm{soc}}^{xy}$ ($H_{\mathrm{soc}}^{z}$) must be the products of $k_\pm=k_x\pm ik_y$ and $\hat{\sigma}_\pm=\hat{\sigma}_x\pm i\hat{\sigma}_y$ ($\hat{\sigma}_z$)  for the invariance,
\begin{align}
H_{\mathrm{soc}}^{xy}=&\begin{pmatrix}
0&0&0\\
0&0&-i{\beta}k_-\hat{\sigma}_-\\
0&i{\beta}k_+\hat{\sigma}_+
\end{pmatrix},\\
H_{\mathrm{soc}}^{z}=&\begin{pmatrix}
0&0&0\\
0&0&-i{\gamma}k_+\hat{\sigma}_z\\
0&i{\gamma}k_-\hat{\sigma}_z
\end{pmatrix},
\end{align}
where the coupling constants $\beta$ and $\gamma$ represent the materials.
Here, no inter-orbital component associated with $|1\rangle$ is included because of the suppression due to the large excitation energy $E_1$ in comparison with that between the conduction and valence bands.
Therefore, the effective Hamiltonian of Janus double-transition-metal MXenes, Mo$_2$HfC$_2$OS and W$_2$HfC$_2$OS, is represented by
\begin{align}
H=H_0+H_{\mathrm{soc}}^{\mathrm{LS}}+H_{\mathrm{soc}}^{\mathrm{R}}+H_{\mathrm{soc}}^{xy}+H_{\mathrm{soc}}^{z},\label{eq_spinful_model}
\end{align}
up to the first order with respect to the wave number $k$.


The effect of the inter-orbital coupling is shown in Fig.\;\ref{fig_effective_bands}(b)-(c).
The Rashba coupling induces the isotropic spin-split increasing with the wave number $k$, but the split does not involve the linear intersection with respect to $k$ unlike that in the band of $p$-orbitals.\cite{Manchon2015}
On the other hand, the inter-orbital coupling to the in-plane spin, $H_{\mathrm{soc}}^{xy}$, splits the band with a linear intersection at the $\Gamma$ point as shown in Fig.\;\ref{fig_effective_bands}(c).
These couplings to the in-plane spin generate the spin-split, but they do not cause the anisotropic behavior unlike the first-principles calculation in Fig.\;\ref{fig_bands}.
However, the inter-orbital coupling to the out-of-plane spin, $H_{\mathrm{soc}}^{z}$, leads to the anisotropic split as shown in Fig.\;\ref{fig_effective_bands}(d).
Then, the spin-split possesses threefold-rotational periodicity around the $\Gamma$ point and it closes along the $\Gamma$-$M$ path. 
Therefore, these three terms cannot be complementary and all the terms are necessary for the effective model to describe the characteristics of the first-principles band structures.


\subsection{Reduced representations near the $\Gamma$ point}


Around the $\Gamma$ point, the spin-orbit coupling can also be approximated by an alternative representation on the basis of $(E_{\mathrm{c}},\;E_{\mathrm{v}})$ equivalent to Eq.\;(\ref{eq_LS_effective}).
The Rashba coupling preserves the diagonal components under the transformation using $\hat{\Psi}_0$ in Eq.\;(\ref{eq_state_matrix}), 
\begin{align}
h^{\mathrm{R}}_{\mathrm{soc}}=-\begin{pmatrix}
\alpha(k_x\hat{\sigma}_y-k_y\hat{\sigma}_x)&0\\
0&\alpha(k_x\hat{\sigma}_y-k_y\hat{\sigma}_x)
\end{pmatrix},
\end{align}
with the limit of $c_r\rightarrow1$ and $d_r\rightarrow0$.
In the same basis, the inter-orbital coupling to the in-plane spin plays the similar role of the Rashba coupling,
\begin{align}
h^{xy}_{\mathrm{soc}}=&\begin{pmatrix}
\beta(k_x\hat{\sigma}_y-k_y\hat{\sigma}_x)&0\\
0&-\beta(k_x\hat{\sigma}_y-k_y\hat{\sigma}_x)
\end{pmatrix}\nonumber\\
&-\begin{pmatrix}
0&-i\beta(k_x\hat{\sigma}_x+k_y\hat{\sigma}_y)\\
i\beta(k_x\hat{\sigma}_x+k_y\hat{\sigma}_y)&0
\end{pmatrix},
\end{align}
but the first term and second term lead to different relative angle between the spin and wave number, the chiral and herical spin configurations, respectively.
The inter-orbital coupling to the out-of-plane spin is transformed into
\begin{align}
h^{z}_{\mathrm{soc}}=&-\begin{pmatrix}
\gamma k\sin3\theta_k\hat{\sigma}_z&0\\
0&-\gamma k\sin3\theta_k\hat{\sigma}_z
\end{pmatrix}\nonumber\\
&-\begin{pmatrix}
0&-i\gamma k\cos3\theta_k\hat{\sigma}_z\\
i\gamma k\cos3\theta_k\hat{\sigma}_z&0
\end{pmatrix}.
\end{align}
Here, the representation is not equivalent to the two-dimensional form of Rashba coupling on the $d$-orbital base\cite{Shanavas2014} or two-dimensional Dresselhaus coupling\cite{Yang2021}.
Then, in this basis, the representation exhibits the contribution to the trigonal variation in the coupling amplitude explicitly.
Although these representations reveal the variation of spin direction with the wave number in each term, they do not well describe the behavior of spin-split in each band.


Although the above approximation enables us to reduce the effective Hamiltonian into the basis of $(E_{\mathrm{c}},\;E_{\mathrm{v}})$, it is not analytically solvable except for $\boldsymbol{k}$ along the $\Gamma$-$K$ path.
However, the behavior of spin-split around the $\Gamma$ point can be analyzed transforming the spin-orbit coupling terms onto the basis of the conduction and valence bands in the presence only of LS coupling, i.e., the spatial parity symmetric basis.
The reduced Hamiltonian of $H_0$ and $H_{\mathrm{soc}}^{\mathrm{LS}}$ is given by
\begin{align}
h_0=\begin{pmatrix}
uk^2\hat{\sigma}_0&\lambda\hat{\sigma}_z\\
\lambda\hat{\sigma}_z&-uk^2\hat{\sigma}_0
\end{pmatrix}.
\end{align}
Since the Hamiltonian can be decomposed to two $2\times2$ blocks for the up- and down-spin along the $z$-axis, two degenerated eigenstates are represented by $\phi^a_\uparrow$ and $\phi^a_\downarrow$ with the spin axis in the conduction band $a=c$ and the valence band $a=v$.
Then, the parity-breaking terms can be rewritten on the basis of $\hat{\phi}_s=[\phi_\uparrow^c,\phi_\downarrow^c,\phi_\uparrow^v,\phi_\downarrow^v]$ with a scaling factor $k_\lambda^2=\lambda/u$ for the wave number.
On this basis, the Rashba coupling is represented with $\zeta_k=\sqrt{1+(k/k_\lambda)^4}$ by
\begin{align}
\hat{\phi}_s^\dagger h_{\mathrm{soc}}^{\mathrm{R}}\hat{\phi}_s=&
-
\frac{\alpha(k/k_\lambda)^2}{\zeta_k}(k_x\hat{\sigma}_y-k_y\hat{\sigma}_x)
\begin{pmatrix}
1&0\\
0&1
\end{pmatrix}
\nonumber\\
&
-\frac{\alpha}{\zeta_k}(k_x\hat{\sigma}_x+k_y\hat{\sigma}_y)\begin{pmatrix}
0&-i\\
i&0\\
\end{pmatrix}\label{eq_Rashba_in_bands}
.
\end{align}
Here, no $k$-linear intra-band term is not included in $h_{\mathrm{soc}}^{\mathrm{R}}$. 
On the other hand, the inter-orbital coupling to the in-plane spin, $h_{\mathrm{soc}}^{xy}$, causes the $k$-linear intra-band coupling,
\begin{align}
\hat{\phi}_s^\dagger h_{\mathrm{soc}}^{xy}\hat{\phi}_s=&\beta(k_x\hat{\sigma}_y-k_y\hat{\sigma}_x)\left\{
\begin{pmatrix}
1&0\\
0&-1\\
\end{pmatrix}
+\frac{1}{\zeta_k}
\begin{pmatrix}
1&0\\
0&1\\
\end{pmatrix}
\right\}
\nonumber\\
&-
\frac{\beta(k/k_\lambda)^2}{\zeta_k}(k_x\hat{\sigma}_x+k_y\hat{\sigma}_y)
\begin{pmatrix}
0&-i\\
i&0
\end{pmatrix},\label{eq_xy_in_bands}
\end{align}
but the intra-band coupling affects only on the conduction band in the vicinity of the $\Gamma$ point because a small wave number leads to $\zeta_k\simeq1$.
This inequivalnce in two bands is consistent with the observation on the first-principles band structures in Fig.\;(\ref{fig_bands}).
The trigonally fluctuating inter-orbital coupling can be represented by
\begin{align}
\hat{\phi}_s^\dagger h_{\mathrm{soc}}^{z}\hat{\phi}_s=&
-\frac{\gamma (k/k_\lambda)^2}{\zeta_k}k \sin3\theta_k\hat{\sigma}_z
\begin{pmatrix}
1&0\\
0&-1
\end{pmatrix}
\nonumber\\
&
+\frac{\gamma}{\zeta_k} k\sin3\theta_k\hat{\sigma}_0
\begin{pmatrix}
0&1\\
1&0\\
\end{pmatrix}
-\gamma k\cos3\theta_k\hat{\sigma}_z
\begin{pmatrix}
0&-i\\
i&0\\
\end{pmatrix}\label{eq_z_in_bands}
,
\end{align}
and explicitly indicates that no $k$-linear intra-band term is included.
However, since $k_\lambda$ is much smaller than the $\pi/a$ practically, the trigonal fluctuation can be observed around the $\Gamma$ point.
Thus, the representations on the basis of $\hat{\psi}_s$ well describe the roles of the inter-orbital spin-orbit coupling to produce the characteristics of the first-principles band structure in the lower panels of Fig.\;(\ref{fig_bands}) (a) and (b). 


\subsection{Model parameters}


\begin{figure}[htbp]
\begin{center}
 \includegraphics[width=80mm]{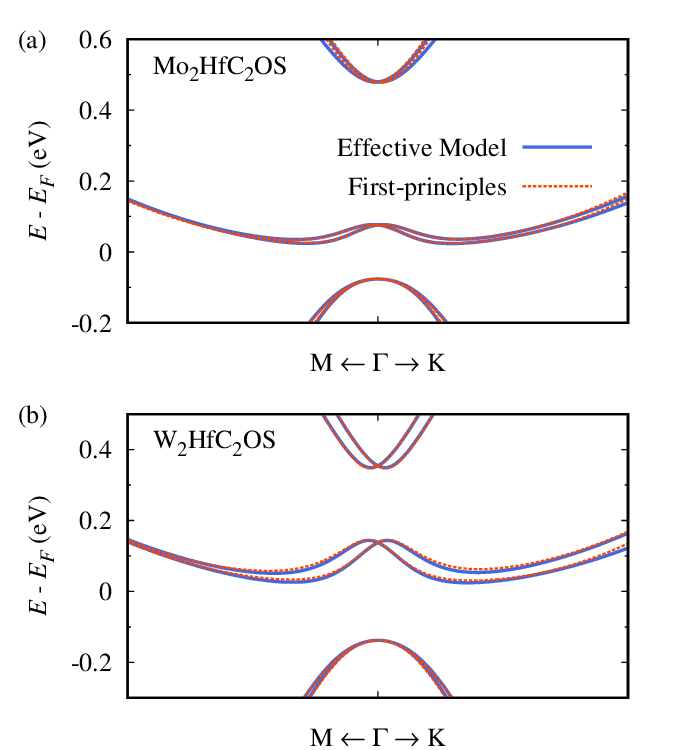}
\caption{The band structures obtained using the effective model in Eq.\;(\ref{eq_spinful_model}) with the parameters in table\ref{tab_model_parameters}. In (a) and (b), those for Mo$_2$HfC$_2$OS and W$_2$HfC$_2$OS (solid-lines) are presented, respectively, with the first-principles bands (dashed-lines).
 }\label{fig_spinful_effective_bands}
\end{center}
\end{figure}


In table \ref{tab_model_parameters}, we provide the parameters for the model Hamiltonian to simulate the first-principles band structure around the $\Gamma$ point.
Moreover, the simulated energy dispersion is presented in Fig.\;\ref{fig_spinful_effective_bands} with the first-principles band structure for Mo$_2$HfC$_2$OS and W$_2$HfC$_2$OS.
The numerical simulation explicitly indicates the reproducibility of the effective model in terms of the characteristics of band structure: the presence and absence of the linear intersection at the $\Gamma$ point, and trigonal anisotropy in the dispersion.
The parameters also enable us not only to calculate the similar energy dispersion but also to investigate the electronic structure in the Janus double-transition-metal MXenes.
In the next section, we discuss the utility of the effective model for the description of the spin configuration as an application of this effective model.


\begin{table}
\caption{The parameters for the effective model to reproduce the first-principles band structures of Mo$_2$HfC$_2$OS and W$_2$HfC$_2$OS. The unit for the energies, $E_1$ and $\lambda$, is eV. The parameters for the $k$-linear and quadratic terms are presented in the unit of eV$\cdot$\AA and eV$\cdot$\AA$^2$, respectively.
}
\begin{ruledtabular}
\begin{tabular}{c c c c c c c c c}
&$E_1$&$u$&$v$&$\lambda$&$\alpha_0$&$\alpha$&$\beta$&$\gamma$\\ \hline
Mo&0.480&2.610&3.790&0.076&-0.290&0.170&0.180&0.040\\ 
W&0.354&2.617&3.906&0.138&-1.500&0.660&0.690&0.090\\ 
\end{tabular}\label{tab_model_parameters}
\end{ruledtabular}
\end{table}


\section{Effect of spin-orbit coupling on the spin configuration}\label{sec_spin_configuration}


\begin{figure}[htbp]
\begin{center}
 \includegraphics[width=80mm]{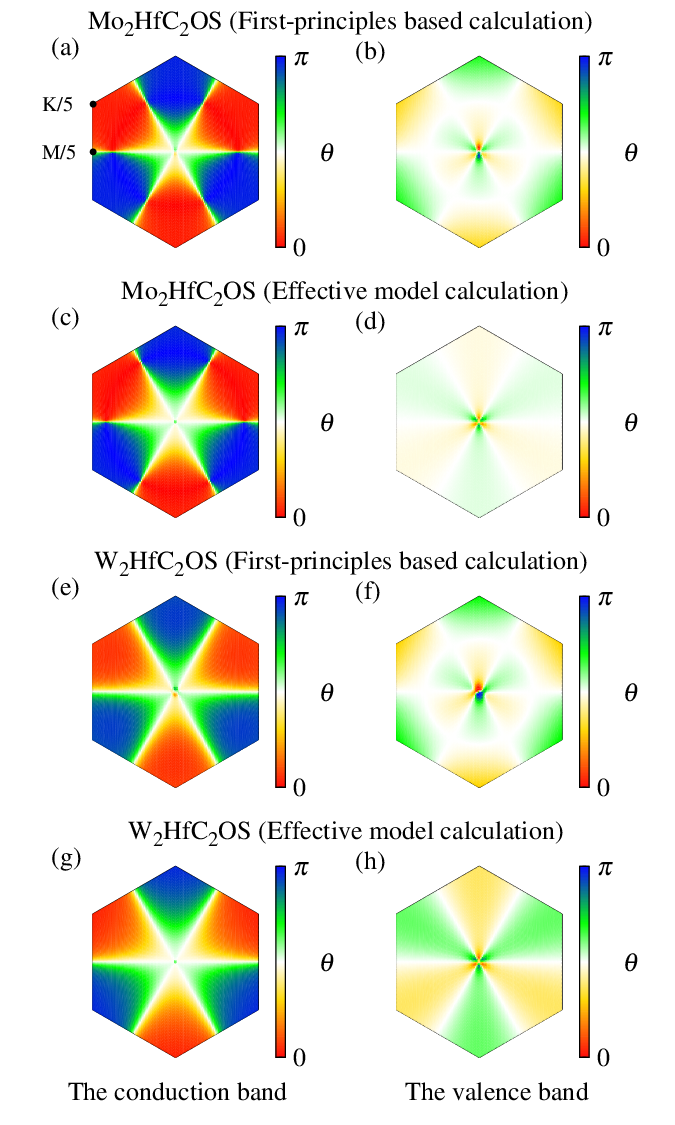}
\caption{The spin angle $\theta$ of each electronic state at $\boldsymbol{k}$ around the $\Gamma$ point with respect to the $z$-axis. Here, $\theta=0$ and $\pi$ indicate the spin is parallel and anti-parallel to the $z$-axis, respectively. The left and right panels show the angle for the conduction and valence bands, respectively.
 }\label{fig_spin_configuration}
\end{center}
\end{figure}


Although the effective model is developed to reproduce the energy dispersion, it can also describe the spin configuration of electronic states in the reciprocal space.
Since each band is split into two spin branches by the spin-orbit coupling, each electronic state possesses a specific spin direction in a single branch.
The spin is not fully polarized in the $z$-axis and it possess the parallel component to the layer, i.e., a tilt with respect to the $z$-axis.
In Fig.\;\ref{fig_spin_configuration}, the spin angle $\theta$ with respect to the $z$-axis is depicted for each electronic state around the $\Gamma$ point for the lower spin branches of the conduction and valence bands.
For Mo$_2$HfC$_2$OS, two different methods are adopted for calculating the spin angle: the first-principles based calculation and the effective model calculation.
For the former calculation, a multi-orbital tight-biding model is generated using Wannier90,\cite{Wannier90} a code to obtain the maximally-localized Wannier functions and the hopping matrix among the orbitals referring to a first-principles band structure.
The tight-binding model is defined on the basis of 54 Wannier orbitals, five $d$-orbitals for transition-metal atoms and three $p$-orbitals for others, including spin degree of freedom, and it well reproduces the corresponding DFT band structure (see Supplemental Material). 
Here, the computational conditions are the same as those for the symmetrically-terminated counterparts in Ref.\onlinecite{Habe2025-1,Habe2025-2}, i.e., the $12\times12\times1$ $k$-mesh, the convergence tolerance $10^{-10}$\AA$^2$, the disentanglement energy window $[-10,10]$ and the frozen window $[-10,3]$ in the unit of eV.
In Fig.\;\ref{fig_spin_configuration}(a)-(h), the different methods provide the consistent results of spin texture around the $\Gamma$ point.
In the conduction band, the consistency is preserved over the portion of the Brillouin zone. 
On the other hand, in the valence band, the discrepancy can be found near the boundary of the region.
This is because the valence band is highly dispersive and approaches the second highest valence band near the boundary.


The theoretical results in Fig.\;\ref{fig_spin_configuration} indicates that the electronic states exhibit a staggered spin polarization along the $z$ axis with a trigonal pattern around the $\Gamma$ point in the reciprocal space.
Near the $\Gamma$ point, the trigonal pattern can be observed, but the spin axis obviously tilts to the $xy$-plane in both the bands.
In the valence band, the spin configuration preserves the tilt, and it is far from fully polarizing along the $z$ direction even with the increase in $k$.
On the other hand, in the conduction band, almost the full polarization can be observed in the same region except for the vicinity of the $\Gamma$ point.
The inequivalence can be described using the effective model of the spin-orbit coupling.
As shown table \ref{tab_model_parameters}, the coefficients $\alpha$ and $\beta$ for the Rashba-type coupling are much larger than $\gamma$ to generate the trigonal pattern of the $z$ spin.
Therefore, each Rashba-type intra-band term, which tilts the spin to the $xy$-plane, are stronger than the coupling to the $z$ spin.
However, the total Rashba-type coupling, $h_{\mathrm{soc}}^{\mathrm{R}}+h_{\mathrm{soc}}^{xy}$, possesses the practical coupling constant, $\beta-\alpha$, smaller than $\gamma$ in the conduction band for $k_\lambda\ll k$, i.e., $\zeta_k\sim(k/k_\lambda)^2$, according to Eqs.\;(\ref{eq_Rashba_in_bands}) and (\ref{eq_xy_in_bands}).
Then, the trigonal spin fluctuation along the $z$-axis dominates the spin configuration in the conduction band.
In the valence band, on the other hand, the practical coupling constant for tilting the spin axis is given by $\alpha+\beta$, which is much larger than $\gamma$.
Thus, the different behavior of spin configuration in the two bands is attributed to the difference of the practical coupling constant for the total Rashba-type coupling as described by the effective model.


The above analysis using the effective model indicates that the magnitude of the coupling constant $\gamma$ does not directly dominate the accuracy of staggered spin polarization along the $z$ axis and that the ratio of coupling constants,
\begin{align}
\rho=\left|\frac{\gamma}{\alpha-\beta}\right|,
\end{align}
is significant for generating the accurate polarization.
This property can be clearly observed in the difference of spin configuration between two compounds, Mo$_2$HfC$_2$OS and W$_2$HfC$_2$OS, in Fig.\;\ref{fig_spin_configuration}(c) and (e).
Tungsten compounds in general possess larger coefficients for the spin-orbit coupling than those of molybdenum compounds because of the larger atomic number.
Thus, the strong spin-orbit coupling is obtained in W$_2$HfC$_2$OS, but the accurate $z$-spin polarization can be achieved in Mo$_2$HfC$_2$OS obviously.
This is because the ratio of coefficient $\rho$ is enhanced in the molybdenum compound as shown in table \ref{tab_model_parameters}.
In general, a heavier element is adopted to obtain more drastic effect in terms of spin configuration, but it is not the proper strategy for the Janus double-transition-metal MXenes.
For these materials, the effective model reveals that the large spin-split and accurate staggered spin polarization are attributed to the magnitudes of $\beta$ and the ratio of coefficients $\rho$, respectively. 
The result means that a large spin-split and an accurate staggered spin polarization are not competed to each other in principle.
Therefore, the effective model clearly indicates that it is possible to realize a material with a large spin-split and an accurate staggered spin polarization along the $z$-axis simultaneously.


\section{discussion}\label{sec_discussion}


In this work, we revealed that the asymmetric termination leads to the unconventional trigonal spin-momentum locking of out-of-plane spin in the double-transition-metal MXenes $M_2$HfC$_2$OS.
The spin configuration possesses a similar property to Rashba systems, i.e., the inverted spin polarization at the opposite wave numbers in a single Fermi pocket, but the polarization axis is almost perpendicular to the layer unlike Rashba systems.
The spin variation in a single Fermi pocket enables us to manipulate the spin of conduction electrons with changing the motion.
Contrarily, transition-metal dichalcogenides also gives a spin-momentum locking of out-of-plane spin, but the polarization direction is uniform over a single Fermi pocket.\cite{Xiao2012}
Therefore, the staggered spin polarization provides the unique system for the manipulation of out-of-plane spin with controlling electronic motion in spintroincs.
Moreover, in the vicinity of the $\Gamma$ point, the Janus double-transition-metal MXenes possess large Rashba couplings from 0.170 to 0.660 eV$\cdot$\AA .
Thus, the isolated two-dimensional bulk materials possess the coupling constant close to a giant Rashba splitting at the surface of a metal\cite{Koroteev2004} or the interface of a heterostructure.\cite{Yi2022}

In the present DFT calculations, a non-hybrid functional is adopted, but it usually underestimates the band gap. 
Actually, Heyd-Scuseria-Ernzerhof functional a hybrid functional,\cite{Heyd2003} improves the direct gap at the $\Gamma$ point from 0.152 eV to 0.209 eV for Mo$_2$HfC$_2$OS and from 0.276 eV to 0.355 eV for W$_2$HfC$_2$OS.
The improvement implies the enhancement of a spin-orbit coupling constant $\lambda$ to 0.105 eV and 0.178 eV for the molybdenum and tungsten compounds, respectively.


\section{conclusion}\label{sec_conclusion}


In this paper, we have developed an effective three-band model to describe the spin-orbit coupling generated by the Janus structure of the double-transition-metal MXenes, $M_2$HfC$_2$OS for $M=$Mo and W.
The effective model reveals that the spin-orbit coupling correlates three degrees of freedom: the spin, atomic orbital, and Bloch dynamics, i.e., the wave number.
Thus, the mathematical representation is not equated with conventional forms, LS, Rashba, and Dresselhaus couplings, even in the reduced one.
The model of spin-orbit coupling enables us to well reproduce the spin-split dispersion in the electronic structure obtained using first-principles calculations near the insulating gap and the spin-configuration of electronic states around the $\Gamma$ point in the reciprocal space.
The spin configuration exhibits a trigonal pattern of staggered spin polarization along the perpendicular direction to the layer plane.
Especially in the conduction band, the electronic spin approaches to be fully polarized in the vertical axis with distance from the $\Gamma$ point.
The effective model also reveals that the enhancement of polarization is attributed to the ratio of coupling constants rather than the magnitude of them.
Thus, the effective representation of the spin-orbit coupling enable us not only to reproduce the unconventional spin-split dispersion in the Janus materials but also to describe the spin structure of electronic states for further qualitative investigations of the effects attributed to the spin-orbit coupling among the spin, atomic orbital, and Bloch dynamics.

\begin{acknowledgements}
This work was supported by JSPS KAKENHI Grant Number JP23K03289.
\end{acknowledgements} 

\bibliography{topological_MXene}

\end{document}